%% file: main.tex
\documentclass[runningheads]{llncs}
\usepackage[T1]{fontenc}
\usepackage{graphicx}
\usepackage{subcaption}
\usepackage{adjustbox}
\usepackage{float}
\usepackage{placeins}
\begin{document}
\title{High-Resolution Cranial Defect Reconstruction by Iterative, Low-Resolution,  Point Cloud Completion Transformers}
\titlerunning{High-Resolution Cranial Defect Reconstruction by Point Cloud Transformers}
\author{Marek Wodzinski\inst{1, 2}\orcidID{0000-0002-8076-6246} \and
Mateusz Daniol\inst{1}\orcidID{0000-0003-2363-7912} \and
Daria Hemmerling\inst{1}\orcidID{0000-0002-2193-7690} \and Miroslaw Socha\inst{1}\orcidID{0000-0001-9462-8269}}
%
\authorrunning{M. Wodzinski et al.}
\institute{
$^{1}$AGH University of Science and Technology \\ Department of Measurement and Electronics, Krakow, Poland \\
$^{2}$University of Applied Sciences Western Switzerland (HES-SO Valais) \\ Information Systems Institute, Sierre, Switzerland \\
\email{
wodzinski@agh.edu.pl \\
}
}
\maketitle              
\begin{center}
\textbf{The preprint has not undergone peer review or any post-submission improvements or corrections. The DOI of Version of Record of this contribution: \url{https://doi.org/10.1007/978-3-031-43996-4_32} and is available online at: \url{https://link.springer.com/chapter/10.1007/978-3-031-43996-4_32}.}
\end{center}

\begin{abstract}
Each year thousands of people suffer from various types of cranial injuries and require personalized implants whose manual design is expensive and time-consuming. Therefore, an automatic, dedicated system to increase the availability of personalized cranial reconstruction is highly desirable.
The problem of the automatic cranial defect reconstruction can be formulated as the shape completion task and solved using dedicated deep networks. Currently, the most common approach is to use the volumetric representation and apply deep networks dedicated to image segmentation.
However, this approach has several limitations and does not scale well into high-resolution volumes, nor takes into account the data sparsity. In our work, we reformulate the problem into a point cloud completion task. We propose an iterative, transformer-based method to reconstruct the cranial defect at any resolution while also being fast and resource-efficient during training and inference. We compare the proposed methods to the state-of-the-art volumetric approaches and show superior performance in terms of GPU memory consumption while maintaining high-quality of the reconstructed defects.

\end{abstract}
\input{introduction}

\input{methods}

\input{results}

\input{discussion}

\FloatBarrier

\section*{Acknowledgements} The project was funded by The National Centre for Research and Development, Poland under Lider Grant no: LIDER13/0038/2022 (DeepImplant). We gratefully acknowledge Polish HPC infrastructure PLGrid support within computational grant no. PLG/2023/016239.

%
%
\bibliographystyle{splncs04}
\bibliography{bibliography}
\end{document}

%% file: introduction.tex
\section{Introduction}

The cranial damages are a common outcome of traffic accidents, neurosurgery, and warfare. Each year, thousands of patients require personalized cranial implants~\cite{global}. Nevertheless, the design and production of personalized implants are expensive and time-consuming. Nowadays, it requires trained employees working with computer-aided design (CAD) software~\cite{design}. However, one part of the design pipeline, namely defect reconstruction, can be directly improved by the use of deep learning algorithms~\cite{autoimplant1,autoimplant2}.

The problem can be formulated as a shape completion task and solved by dedicated neural networks. Its importance motivated researchers to organize two editions of the AutoImplant challenge, during which researchers proposed several unique contributions~\cite{autoimplant1,autoimplant2}. The winning contributions from the first~\cite{ellis1} and second editions~\cite{wodzinski1} proposed heavily-augmented U-Net-based networks and treated the problem as segmentation of missing skull fragment. They have shown that data augmentation is crucial to obtain reasonable results. Other researchers proposed similar encoder-decoder approaches, however, without significant augmentation and thus limited performance~\cite{mahdi,pathak}. Another group of contributions attempted to address not only the raw performance but also the computational efficiency and hardware requirements. One contribution proposed an RNN-based approach using 2-D slices taking into account adjacent slices to enforce the continuity of the segmentation mask~\cite{yang}. The contribution by Li \textit{et al.} has taken into account the data sparsity and proposed a method for voxel rearrangement in coarse representation using the high-resolution templates~\cite{li1}. The method was able to substantially reduce memory usage while maintaining reasonable results. Another contribution by Kroviakov \textit{et al.} proposed an approach based on sparse convolutional neural networks~\cite{kroviakov1} using Minkowski engine~\cite{choy}. The method excluded the empty voxels from the input volume and decreased the number of the required convolutions. The work by Yu \textit{et al.} proposed an approach based on principal component analysis with great generalizability, yet limited raw performance~\cite{yu}. Interestingly, methods addressing the computational efficiency could not compete, in terms of the reconstruction quality, with the resource-inefficient methods using dense volumetric representation~\cite{autoimplant2}.

The current state-of-the-art solutions, even though they reconstruct the defects accurately, share some common disadvantages. First, they operate in the volumetric domain and require significant computational resources. 
The GPU memory consumption scales cubically with the volume size. Second, the most successful solutions do not take into account data sparsity. The segmented skulls are binary and occupy only a limited part of the input volume. 
Thus, using methods dedicated to 3-D multi-channel volumes is resource-inefficient. Third, the final goal of the defect reconstruction is to propose models ready for 3-D printing. Working with volumetric representation requires further postprocessing to transfer the reconstructed defect into a printable model.

Another approach, yet still unexplored, to cranial defect reconstruction is the use of deep networks dedicated to point clouds (PCs) processing. Since the introduction of PointNet~\cite{pointnet} and PointNet++\cite{pointnet++}, the number of contributions in the area of deep learning for PC processing exploded. Several notable contributions, like PCNet~\cite{pcnet}, PoinTr~\cite{pointr}, AdaPoinTr~\cite{adapointr}, 3DSGrasp~\cite{3dgrasp}, MaS~\cite{mas}, have been proposed directly to the PC completion task. The goal of the PC completion is to predict a missing part of an incomplete PC. 

The problem of cranial defect reconstruction can be reformulated into PC completion which has several advantages. First, the representation is sparse, and thus requires significantly less memory than the volumetric one. Second, PCs are unordered collections and can be easily splitted and combined, enabling further optimizations. Nevertheless, the current PCs completion methods focus mostly on data representing object surfaces and do not explore large-scale PCs representing solid objects.

In this work, we reformulate the problem from volumetric segmentation into PC completion. We propose a dedicated method to complete large-scale PCs representing solid objects. We extend the geometric aware transformers~\cite{pointr} and propose an iterative pipeline to maintain low memory consumption. We compare the proposed approach to the state-of-the-art networks for volumetric segmentation and PC completion. Our approach provides high-quality reconstructions while maintaining computational efficiency 
and good generalizability into previously unseen cases.

%% file: methods.tex
\section{Methods}

\subsection{Overview}

\begin{figure*}[!htb]
	\centering
    \includegraphics[scale=0.60]{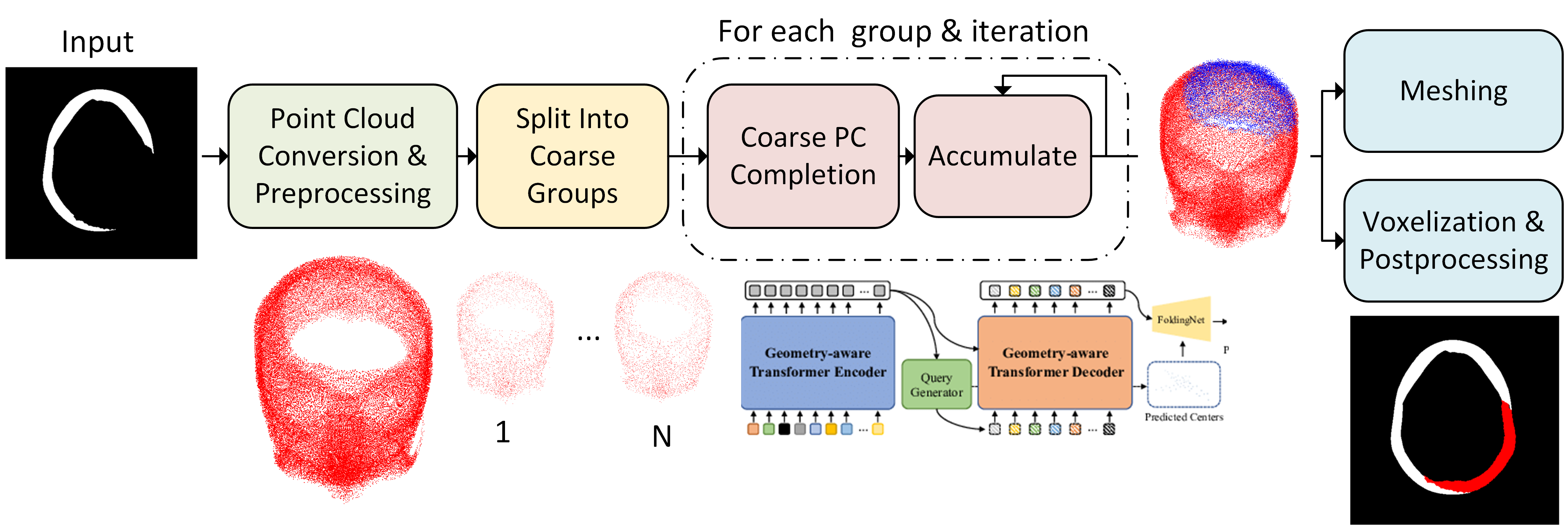}
    \caption{Visualization of the processing pipeline.}   
    \label{fig:pipeline}
\end{figure*}

The input is a 3-D binary volume representing the defective skull. The output is a PC representing the missing skull fragment and (optionally) its meshed and voxelized representation. The processing pipeline consists of: (i) creating the PC from the binary volume, (ii) splitting the PC into a group of coarse PCs, (iii) calculating the missing PC by the geometric aware transformer for each group, (iv) merging the reconstructed coarse PCs, (v) optional voxelization and postprocessing for evaluation. The pipeline is shown in Figure~\ref{fig:pipeline}.

\subsection{Preprocessing}

The preprocessing starts with converting the binary volume to the PC. The coordinates of the positive voxels are created only from the voxels representing the skull. The PC is normalized to [0-1] range, randomly permuted, and split into $N$ equal groups, where $N$ is calculated based on the number of points in the input PC in a manner that each group contains 32768 points and outputs 16384 points.

\subsection{Network Architecture - Point Cloud Completion Transformer}

We adapt and modify the geometry-aware transformers (PoinTr)~\cite{pointr}. The PoinTr method was proposed and evaluated on coarse PCs representing object surfaces. The full description of the PoinTr architecture is available in~\cite{pointr}.


We modify the network by replacing the FoldingNet~\cite{foldingnet} decoder working on 2-D grids with a folding decoder operating on 3-D representation. The original formulation deforms the 2-D grid into the surface of a 3-D object, while the proposed method focuses on solid 3-D models. Moreover, we modify the original k-NN implementation (with quadratic growth of memory consumption with respect to the input size) to an iterative one, to further decrease and stabilize the GPU memory consumption.

\subsection{Objective Function}

We train the network supervisedly where the ground-truth is represented by PCs created from the skull defects. In contrast to other PC completion methods, we employ the Density Aware Chamfer Distance (DACD)~\cite{densitychamfer}. The objective function enforces the uniform density of the output and handles the unpredictable ratio between the input/output PCs size. We further extend the DACD by calculating the distance between the nearest neighbours for each point and enforcing the distance to be equal. The final objective function is:
\begin{equation}
    O(P_{r}, P_{gt}) = DACD(P_{r}, P_{gt}) +\frac{\alpha}{S}\sum_{i=0}^S\sum_{j=0}^k\sum_{l=0}^k|P_{r}(i) - P_{r}(j)| - |P_{r}(i) - P_{r}(l)|,
\end{equation}
where $P_{r}, P_{gt}$ are the reconstructed and ground-truth PC respectively, $S$ is the number of points in $P_{rec}$, $k$ is the number of nearest neighbours of point $i$, $\alpha$ is the weighting parameter. We apply the objective function to all PC ablation studies unless explicitly stated otherwise. The volumetric ablation studies use the soft Dice score.

The traditional objective functions like Chamfer Distance (CD)~\cite{densitychamfer}, Extended Chamfer Distance (ECD)~\cite{foldingnet}, or Earth Mover's Distance (EMD)~\cite{mas} are not well suited for the discussed application. The CD/ECD provide suboptimal performance for point clouds with uniform density or a substantially different number of samples, tends to collapse, and results in noisy training. The EMD is more stable, however, explicitly assumes bijective mapping (requiring knowledge about the desired number of points) and has high computational complexity.




\subsection{Iterative Completion}

The coarse PCs are processed by the network separately. Afterwards, the reconstructed PCs are combined into the final reconstruction. To improve the results, the process may be repeated $M$ times with a different initial PC split and a small Gaussian noise added. The procedure improves the method's performance and closes empty holes in the voxelized representation. The optional multi-step completion is performed only during the inference.

The iterative completion allows one to significantly reduce the GPU memory usage and the number of network parameters. The PCs are unordered collections and can be easily split and merged. There is no need to process large PCs in one shot, resulting in the linear growth of inference time and almost constant GPU memory consumption. 

\subsection{Postprocessing}

The reconstructed PCs are converted to mesh and voxelized back to the volumetric representation, mainly for evaluation purposes. The mesh is created by a rolling ball pivoting algorithm using the Open3D library~\cite{open3d}. The voxelization is also performed using Open3D by the PC renormalization and assigning positive values to voxels containing points in their interior. The voxelized representation is further postprocessed by binary closing and connected component analysis to choose only the largest volume. Then, the overlap area between the reconstructed defect and the defective input is subtracted from the reconstructed defect by logical operations. 

\subsection{Dataset and Experimental Setup}

We use the SkullBreak and SkullFix datasets~\cite{dataset} for evaluation. The datasets were used during the AutoImplant I and II challenges and enable comparison to other reconstruction algorithms. The SkullBreak dataset contains 114 high-resolution skulls for training and 20 skulls for testing, each with 5 accompanying defects from various classes, resulting in 570 training and 100 testing cases. All volumes in the SkullBreak dataset are 512 x 512 x 512. The SkullFix dataset is represented by 100 training cases mostly located in the back of the skull with a similar appearance, and additional 110 testing cases. 
The volumes in the SkullFix dataset are 512 x 512 x Z where Z is the number of axial slices. The SkullBreak provides more heterogeneity while the SkullFix is better explored and enables direct comparison to other methods.


We perform several ablation studies. We check the influence of the input physical spacing on the reconstruction quality, training time, and GPU memory consumption. Moreover, we check the generalizability by measuring the gap between the results on the training and the external testing set for each method. We compare our method to the methods dedicated to PC completion: (i) PCNet~\cite{pcnet}, (ii) PoinTr~\cite{pointr}, (iii) AdaPoinTr~\cite{adapointr}, as well as to methods dedicated to volumetric defect reconstruction: (i) 3-D VNet, and (ii) 3-D Residual U-Net. Moreover, we compare the reconstruction quality to results reported by other state-of-the-art methods.

We trained our network separately on the SkullBreak and SkullFix datasets. 
The results are reported for the external test set containing 100 cases for SkullBreak and 110 cases for the SkullFix datasets, the same as in the methods used for comparison. The models are implemented in PyTorch~\cite{pytorch}, trained using a single RTX GeForce 3090. We augment the input PCs by random permutation, cropping, rotation, and translation. The volumetric ablation studies use random rotation and translation with the same parameters as for the PCs. All the methods were trained until convergence. The hyperparameters are reported in the associated repository.

%% file: results.tex
\section{Results}

\begin{figure*}[!htb]
	\centering
    \includegraphics[scale=0.62]{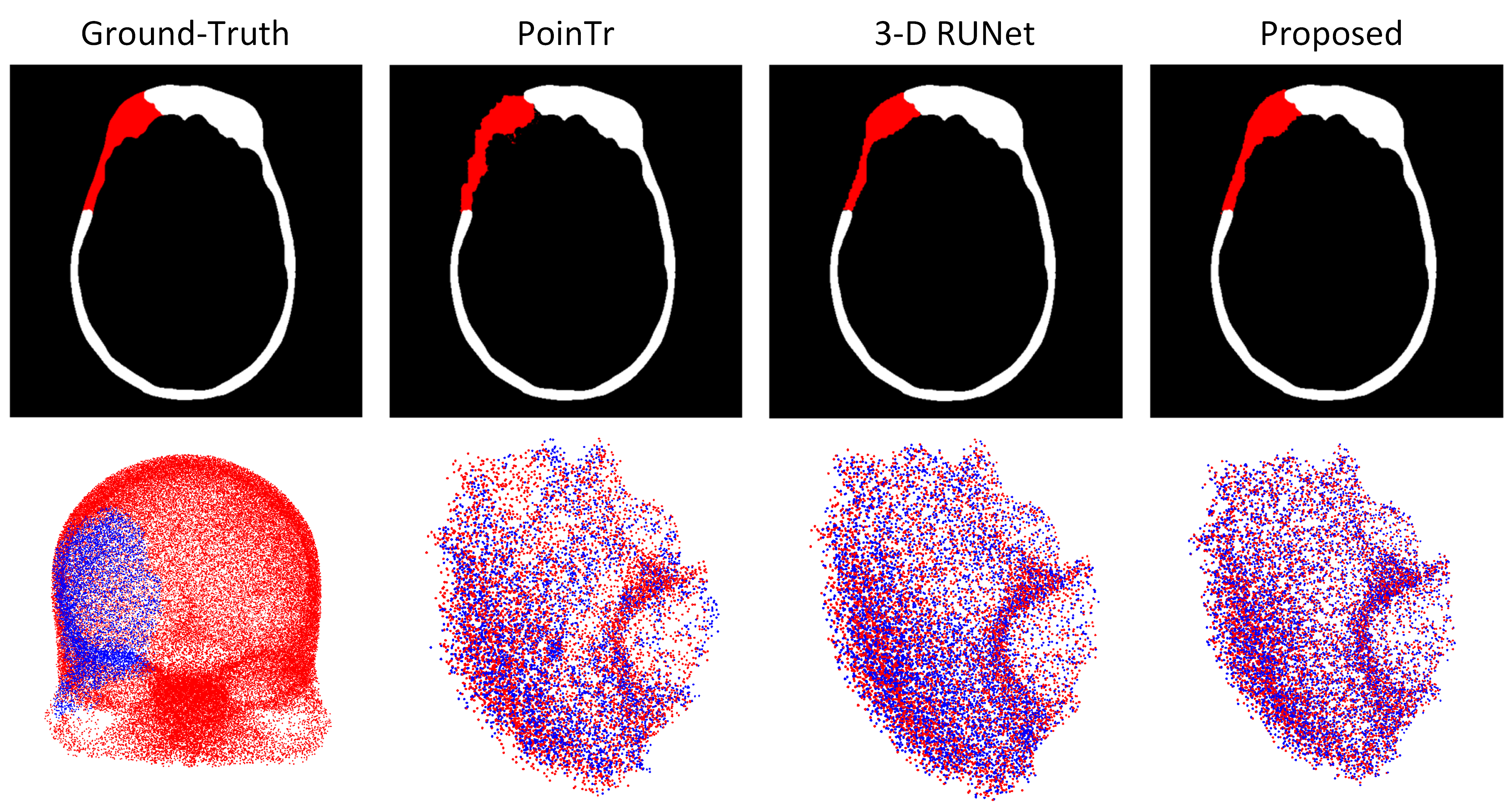}
    \caption{Exemplary visualization of the reconstructed point clouds / volumes for a case from the SkullBreak dataset. The PCs are shown for the defect only (reconstructed vs ground-truth) for the presentation clarity.}   
    \label{fig:visualization}
\end{figure*}

The comparison in terms of the Dice coefficient (DSC), boundary Dice coefficient (BDSC), 95th percentile of Hausdorff distance (HD95), and Chamfer distance (CD), are shown in Table~\ref{tab:results_sb}. Exemplary visualization, presenting both the PC and volumetric outcomes, is shown in Figure~\ref{fig:visualization}.

\begin{table*}[!htb]
\centering
\caption{Quantitative results on the SkullBreak and SkullFix datasets. The final results are reported for original resolution using the DACD + kNN objective function and 3 iterative refinements (- denotes that results were not reported). The methods used for comparison are reported for the most successful setup (see~ Table~\ref{tab:ablation}).}
\renewcommand{\arraystretch}{1.0}
\footnotesize
\resizebox{0.99\textwidth}{!}{%
\begin{tabular}{lccccccccc}
\label{tab:results_sb}
Method &  \multicolumn{4}{c}{SkullBreak} & \multicolumn{4}{c}{SkullFix} & GPU Mem [GB] 
\tabularnewline
 & DSC & BDSC & HD95 & CD &  DSC & BDSC & HD95 & CD 
\tabularnewline
\hline
\multicolumn{10}{c}{Point Cloud Completion} \tabularnewline
\hline
Proposed & 0.87 & 0.85 & 1.91 & 0.31 & 0.90 & 0.89 & 1.71 & 0.29 & $\sim$2.78  \tabularnewline
PCNet~\cite{pcnet} & 0.61 & 0.58 & 5.77 & 1.18 & 0.77 & 0.75 & 3.22 & 0.41 & $\sim$2.37  \tabularnewline
PointTr~\cite{pointr} & 0.67 & 0.66 & 5.17 & 0.82 & 0.82 & 0.81 & 3.02 & 0.36 & $\sim$3.11 \tabularnewline
AdaPoinTr~\cite{adapointr} & 0.66 & 0.64 & 5.29 & 0.84 & 0.81 & 0.81 & 3.05 & -0.36 & $\sim$3.14  \tabularnewline
\hline
\multicolumn{10}{c}{Volumetric Segmentation} \tabularnewline
\hline
3-D VNet & 0.87 & 0.90 & 1.87 & 0.21 & 0.91 & 0.93 & 1.66 & 0.11 & 21.89  \tabularnewline
3-D RUNet & 0.89 & 0.91 & 1.79 & 0.18 & 0.91 & 0.92 & 1.67 & 0.09 & 22.47  \tabularnewline
\hline
\multicolumn{10}{c}{State-of-the-art} \tabularnewline
\hline
Kroviakov \textit{et al.}~\cite{kroviakov1} & - & - & - & - & 0.85 & 0.94 & 2.65 & - & < 6.00 \tabularnewline
Li \textit{et al.}~\cite{li1} & - & - & - & - & 0.81 & - & - & - & -  \tabularnewline
Mahdi \textit{et al.}~\cite{mahdi} & 0.78 & 0.81 & 3.42 & - & 0.88 & 0.92 & 3.59 & - & < 6.00 \tabularnewline
Pathak \textit{et al.}~\cite{pathak} & - & - & - & - & 0.90 & 0.95 & 2.02 & - & -  \tabularnewline
Wodzinski \textit{et al.}~\cite{wodzinski1} & 0.91 & 0.95 & 1.60 & - & 0.93 & 0.95 & 1.48 & - & < 40.00  \tabularnewline
Yu \textit{et al.}~\cite{yu} & - & - & - & - & 0.77 & 0.77 & 3.68 & - & CPU \tabularnewline
Ellis \textit{et al.}~\cite{ellis1} & - & - & - & - & 0.94 & - & 3.60 & - & -  \tabularnewline
Yang \textit{et al.}~\cite{yang} & 0.85 & 0.89 & 3.52 & - & - & - & - & - & -  \tabularnewline
\hline
\end{tabular}}
\end{table*}

The results of the ablation studies showing the influence of the input size, generalizability, objective function, and the effect of repeating the iterative refinement are presented in Table~\ref{tab:ablation}. 

\begin{table*}[!htb]
\centering
\caption{The ablation studies related to the input size, the objective function, and the number of refinements. The results are reported for the SkullBreak dataset at the original scale (except the CD). The Gen. Gap denotes the difference between the training and testing set in terms of the DSC.}
\renewcommand{\arraystretch}{1.0}
\footnotesize
\resizebox{0.99\textwidth}{!}{%
\begin{tabular}{lcccccc}
\label{tab:ablation}
Method & DSC & BDSC & HD95 [mm] & CD [mm] & GPU Mem [GB] & Gen. Gap [\% DSC]
\tabularnewline
\hline 
\multicolumn{7}{c}{Input Size (uniform voxel spacing)} \tabularnewline
\hline
Proposed: original & 0.87 & 0.85 & 1.91 & 0.31 & $\sim$2.78 & 4.18 \tabularnewline
Proposed: 1 mm & 0.83 & 0.77 & 2.64 & 0.46 & $\sim$2.69 & 4.05 \tabularnewline
Proposed: 2 mm & 0.74 & 0.71 & 3.89 & 0.67 & $\sim$2.64 & 4.57 \tabularnewline
Proposed: 4 mm & 0.69 & 0.64 & 5.12 & 0.79 & $\sim$2.63 & 3.12 \tabularnewline
\hline
PCNet: original & - & - & - & - & > 24 & - \tabularnewline
PCNet: 1 mm & 0.37 & 0.33 & 10.57 & 1.76 & $\sim$13.22 & 1.13 \tabularnewline
PCNet: 2 mm & 0.57 & 0.53 & 7.18 & 1.37 & $\sim$5.37 & 1.33 \tabularnewline
PCNet: 4 mm & 0.61 & 0.58 & 5.77 & 1.18 & $\sim$2.37 & 3.07 \tabularnewline
\hline
PoinTr: original & - & - & - & - & > 24 & - \tabularnewline
PoinTr: 1 mm & 0.58 & 0.55 & 6.82 & 1.39 & $\sim$21.41 & 1.89 \tabularnewline
PoinTr: 2 mm & 0.65 & 0.64 & 5.28 & 0.94 & $\sim$6.48 & 2.19 \tabularnewline
PoinTr: 4 mm & 0.67 & 0.66 & 5.17 & 0.82 & $\sim$3.11 & 3.98 \tabularnewline
\hline
3-D RUNet: original & - & - & - & - & > 24 & - \tabularnewline
3-D RUNet: 1 mm & 0.89 & 0.91 & 1.79 & 0.18 & 22.47 & 10.11 \tabularnewline
3-D RUNet: 2 mm & 0.85 & 0.85 & 2.09 & 0.25 & 7.84 & 14.51 \tabularnewline
3-D RUNet: 4 mm & 0.76 & 0.77 & 2.89 & 0.63 & 3.78 & 17.48 \tabularnewline
\hline
\multicolumn{7}{c}{Objective Function (proposed method, original size, 3 iters)} \tabularnewline
\hline
DACD + kNN & 0.87 & 0.85 & 1.91 & 0.31 & $\sim$2.78 & 4.18 \tabularnewline
DACD & 0.85 & 0.81 & 2.78 & 0.42 & $\sim$2.72 & 4.22 \tabularnewline
ECD & 0.75 & 0.71 & 4.11 & 0.68 & $\sim$2.72 & 3.99 \tabularnewline
CD & 0.83 & 0.78 & 2.98 & 0.28 & $\sim$2.72 & 4.58 \tabularnewline
\hline
\multicolumn{7}{c}{No. Refinements (proposed method, original size, DACD + kNN)} \tabularnewline
\hline
1 iters & 0.85 & 0.81 & 2.51 & 0.41 & $\sim$2.78 & 4.51 \tabularnewline
2 iters & 0.87 & 0.83 & 1.98 & 0.32 & $\sim$2.78 & 4.18 \tabularnewline
3 iters & 0.87 & 0.85 & 1.91 & 0.31 & $\sim$2.78 & 4.18 \tabularnewline
4 iters & 0.87 & 0.85 & 1.90 & 0.30 & $\sim$2.78 & 4.18 \tabularnewline
\hline
\end{tabular}}
\end{table*}

%% file: discussion.tex
\section{Discussion}

The reconstruction quality of the method is comparable to the volumetric networks, as shown in Table~\ref{tab:results_sb}. Meanwhile, the proposed method takes into account the data sparsity, does not require significant computational resources, and scales well with the input size. 
The proposed method has good generalizability. The gap between the training and testing set is negligible, unlike the volumetric methods that easily overfit and require strong augmentation for practical use.
The DACD, as well as the proposed extension, improve the reconstruction quality compared to the CD or ECD by taking into account the uniformity of the expected PC. 
The original PC completion methods do not scale well with the increase of PC size. The possible reason for this is connected with the noisy kNN graph construction when dealing with large PCs and increasing the number of neighbours is unacceptable from the computational point of view.
The proposed method has almost constant memory usage, independent of the input shape, in contrast to both the volumetric methods and PC completion methods without the iterative approach.
Interestingly, the proposed method outperforms other methods taking into account the data sparsity. 
The inference speed is slightly lower than for the volumetric methods, however, this application does not require real-time processing and anything in the range of seconds is acceptable.

The disadvantages of the proposed algorithm are connected to long training time, noise at the object boundaries, and holes in the voxelized output. The FoldingNet-based decoder requires a significant number of iterations to converge, thus resulting in training time comparable or even longer than the volumetric methods. Moreover, the voxelization of PCs results in noisy edges and holes that require further morphological postprocessing.

In future work, we plan to further reformulate the problem and, similarly to Kroviakov \textit{et al.}~\cite{kroviakov1}, use only the skull contours. Since the ultimate goal is to propose models ready for 3-D printing, the interior of the skull defect is not required to create the mesh and STL file. 
Another research direction is connected to the PC augmentation to further increase the network generalizability since it was shown that heavy augmentation is crucial to obtain competitive results~\cite{ellis1,wodzinski1}.

To conclude, we proposed a method for cranial defect reconstruction by formulating the problem as the PC completion task. The proposed algorithm achieves comparable results to the best-performing volumetric methods while requiring significantly less computational resources. We plan to further optimize the model by working directly at the skull contour and heavily augmenting the PCs.